\documentclass[twocolumn]{aa}
\usepackage{aalongtable}
\usepackage{graphicx}
\usepackage{amssymb}
\usepackage{natbib}
\bibpunct{(}{)}{;}{a}{}{,}
\usepackage{supertabular}
\usepackage{rotating}
\usepackage{setspace}
\usepackage{lscape}
\usepackage{psfig}
\usepackage{color}
\def\vsini{$V\!\sin i$}

\def\omc{$\Omega/\Omega_{\rm{c}}$}

\def\kps{km~s$^{-1}$}

\def\kms{km~s$^{-1}$}

\begin{document}

\title{Can massive Be/Oe stars be progenitors of long gamma ray bursts?}

\titlerunning{SMC Be/Oe stars as LGRBs progenitors.}
\author{
Christophe Martayan \inst{1,2}
\and  Juan Zorec  \inst{3}
\and Yves Fr\'emat \inst{4}
\and  Sylvia Ekstr\"om  \inst{5}
}
\offprints {C. Martayan}
\mail{Christophe.Martayan@eso.org}

\institute{European Organization for Astronomical Research in the Southern 
Hemisphere, Alonso de Cordova 3107, Vitacura, Santiago de Chile, Chile
\and GEPI, Observatoire de Paris, CNRS, Universit\'e Paris Diderot, 5 place 
Jules Janssen, 92195 Meudon Cedex, France
\and Institut d'Astrophysique de Paris, UMR7095, CNRS, Universit\'e Marie \& Pierre Curie, 
98bis Boulevard Arago 75014 Paris, France
\and Royal Observatory of Belgium, 3 avenue circulaire, 1180 Brussels, Belgium 
\and Geneva Observatory, University of Geneva, Maillettes 51, 1290 Sauverny, Switzerland}

\date{Received / Accepted}
\abstract
{The identification of long-gamma-ray-bursts (LGRBs) is still uncertain, although the collapsar engine of
fast-rotating massive stars is gaining a strong consensus.}
{We propose that low-metallicity Be and Oe stars, which are massive fast rotators, as potential LGRBs
progenitors.}  
{We checked this hypothesis by 1) testing the global specific angular momentum of Oe/Be stars in the ZAMS
with the SMC metallicity, 2) comparing the ZAMS ($\Omega/\Omega_{\rm c},M/M_{\odot}$) parameters of these
stars with the area predicted theoretically for progenitors with metallicity $Z\!=\!0.002$, and 3)
calculating the expected rate of LGRBs/year/galaxy and comparing them with the observed ones. To this end, we
determined the ZAMS linear and angular rotational velocities for SMC Be and Oe stars using the observed
\vsini~ parameters, corrected from the underestimation induced by the gravitational darkening effect.} 
{The angular velocities of SMC Oe/Be stars are on average $\langle\Omega/\Omega_{\rm c}\rangle\!=\!0.95$ in
the ZAMS. These velocities are in the area theoretically predicted for the LGRBs progenitors. We estimated
the yearly rate per galaxy of LGRBs and the number of LGRBs produced in the local Universe up to z=0.2. We
have considered that the mass range of LGRB progenitors corresponds to stars hotter than spectral types B0-B1
and used individual beaming angles from 5 to 15\degr. We thus obtain $R^{\rm pred}_{\rm
LGRB}\sim10^{-7}$ to $\sim10^{-6}$ LGRBs/year/galaxy, which represents on average 2 to 14 LGRB predicted
events in the local Universe during the past 11 years. The predicted rates could widely surpass the observed
ones [(0.2-3)$\times10^{-7}$ LGRBs/year/galaxy; 8 LGRBs observed in the local Universe during the last 11
years] if the stellar counts were made from the spectral type B1-B2, in accordance with the expected
apparent spectral types of the appropriate massive fast rotators.}
{We conclude that the massive Be/Oe stars with SMC metallicity could be LGRBs progenitors. Nevertheless,
other SMC O/B stars without emission lines, which have high enough specific angular momentum, can enhance the
predicted $R_{\rm LGRB}$ rate.} 
\keywords{Gamma rays: bursts -- Stars: early-type -- Stars: emission-line, Be -- Stars: fundamental
parameters --  Galaxies: Magellanic Clouds}

\maketitle

\section{Introduction}

  Since their discovery, data on gamma ray bursts (GRBs) considerably increased, which today allow us to aim at preliminary statistical conclusions. The properties of GRBs and their possible connection to SN were reviewed by \citet{woosley2006} and \citet{fryer2007}. Thus, there are at least 2 classes of GRBs according to the duration of the
phenomenon: 1) short bursts that last less than 1 s and 2) long bursts, which are longer than 1-3 s hereafter LGRBs. Although the short GRBs might correspond to a violent merging of two compact objects, evidence is growing that suggests that the disruption of a massive star can be behind an LGRB \citep{fryer2007}. 
 A possible, but rare, third class of GRBs can be explained by a binary-star scenario \citep{tutu2007}. However,  proper identification of both the progenitor and the final nature of the phenomenon are still fairly uncertain. Nevertheless, a kind of consensus is gaining in the community that the collapsar engine during a supernova SNIb,c explosion, as proposed by \citet{woosley1993}, might lead to allow the explanation of LGRBs. According to this hypothesis, the explosion follows a massive star collapse to a black hole. The circumstellar disk accretes onto the black hole and a bi-polar jet is
formed \citep{hirschi2005}. The infalling material must have enough angular momentum to remain in a disc before accretion.\par  New observational findings and the latest theoretical developments are steadily giving best founded constraints to understand the LGRBs phenomenon. The observations carried out by \citet{iwamoto1998,iwamoto2000} support the idea that massive fast-rotating stars are at the origin of the LGRBs. \citet{thone2008} find that the LGRB \object{GRB060505} is hosted in a low-metallicity galaxy, which is characterized by a high star-formation rate. The event seems to come from a young environment (6 Myrs) and from an object about 32 M$_{\odot}$. The latest models have tried to reproduce the evolution of stars until the end of their lives. In particular, for massive stars, they have been worked out until the GRBs phase. In the search for SNIb,c progenitors, WR stars with He-rich envelopes are recognized as possibly behind the GRBs events. In fact, \citet{hammer2006} find that the LGRBs occur in areas of galaxies with WR stars.\par
  Rotation has been recognized as a key point for understanding the appearance of GRBs \citep{woosley1993,hirschi2005,yoon2006}. Accordingly, to keep a large amount of angular momentum up to the last evolutionary phases before the collapse, GRBs progenitors should be massive objects with low initial metallicities and possibly display anisotropic winds \citep{meynet2007}. Thus, from \citet{yoon2006} it seems that WR stars with metallicities $Z\!\lesssim0.002$ can be progenitors of GRBs. In the same sense, \citet{modjaz2008} find that the SNIc-GRB association occurs when metallicities
are lower than $0.6\times{Z}_{\odot}$. Because stars with low metallicities can on average rotate faster, \citet{hirschi2005} and \citet{yoon2006} foresee that the number of GRBs must grow as the redshift increases, which seems to be confirmed
observationally \citep{fryer2007}. It might still be that the first massive stars, which were very metal-poor stars, were fast rotators. This is an additional reason for the frequency of GRBs being higher as the redshift increases. In this sense, 
\citet{kewley2007} find a link between the location of cosmological LGRBs and the very low-metallicity galaxies.\par
 Thanks to fast rotation and the concomitant efficient mixing of chemical elements, massive stars can undergo quasi-chemically homogeneous evolution to end up as helium WR stars satisfying the requirements for the collapsar scenario 
\citep{yoon2006,vmarle2008}. \citet{yoon2006} have calculated the quasi-chemically homogeneous evolution of magnetized massive stars. They produced diagrams of LGRBs progenitors as a function of their ZAMS rotational velocities and masses and of different initial metallicities. From these diagrams, it emerges that there must be an upper limit to the initial metallicity of LGRBs progenitors, which approximatively corresponds to that of the Small Magellanic Clouds (SMC). According to \citet[][and references therein]{maeder1999}, the SMC average metallicity is $Z\!\sim0.002$. The diagram of \citep{yoon2006} also indicates that, for metallicities $Z\!\lesssim0.002$, the WR phenomenon can appear in stars having lower masses than those in the Milky Way (hereafter MW). In this context, \citet{martins2009} have observed several SMC WR
stars, whose evolutionary status and chemical properties can be understood if they are fast rotators.\par According to the listed observational and theoretically inferred requirements, stars might be potential LGRBs progenitors if they\par

\begin{itemize}
\item are massive enough at the end of their evolution to form a black hole;
\item have lost their hydrogen envelope and have a fast-rotating core; 
\item are formed in low-metallicity environments, where there must also be high star-formation rates to ensure having enough massive stars.
\end{itemize}

 Since the lower the metallicity the faster the rotation and the lower the mass of stars that can undergo the WR phase, according to \citet{yoon2006}, the most massive B-type stars, as well as O-type stars, could become WR stars if they rotated fast enough. The required rotational velocities must be close to those what are typical of Oe/Be stars. Then, we simply ask whether the more massive stars  displaying the Be phenomenon today in the SMC, can be potential progenitors of LGRBs. This possibility has already been raised by \citet{woohe2006}. We recall that the ``Be phenomenon'' appears in the main sequence evolutionary phase of fast-rotating O- and B-type stars. They are characterized by emission lines produced in a decretion disk formed by continuous and episodic matter ejections from the central star. These objects are the fastest known rotators in the main sequence, and their rotation can be closer to the critical one when the metallicity is lower \citep{dds2003,meilland2007,fremat2005,vinicius2006,marta2007}.\par   
 Our research on whether the SMC Oe/Be stars can be progenitors of LGRBs is based on three tests: 1) determination of the global specific stellar angular momentum
(Sect.~\ref{iozrv}), 2) inference of the ZAMS angular velocity ratios of the studied SMC
Oe/Be stars and comparison with the model predicted requirements to be LGRB-progenitor (Sect.~\ref{lgrbp}), and 3) estimation of the LGRBs rate based on the SMC Oe/Be- and fast-rotator population, and its comparison with the observed ones 
(Sect.~\ref{prates}). A general discussion is given in Sect.~\ref{opprd}.\par

\section{Inference of the ZAMS rotational velocities}
\label{iozrv}

\subsection{Main sequence linear equatorial rotational velocities}

  The results presented in this section are based on the FLAMES-GIRAFFE \citep{pasquini2002} observations discussed in \citet{marta2007}\footnote{Based on observations at the European Southern Observatory,  Chile under project number
072.D-0245(A) and (C).}. To the previous sample of 131 Oe/Be stars analyzed by \citet{marta2007}, we added here the most massive Be stars and Oe stars found in the SMC. The projected rotational velocity $V\!\sin i$ of each star was determined with the GIRFIT code \citep{fremat2006}. The average values of $V\!\sin i$ by mass category thus obtained are given in Table~\ref{table1}. We know, however, that Oe/Be stars are fast rotators, and the gravitational darkening effect (hereafter GD) induces systematic underestimations of the $V\!\sin i$ parameters \citep{town2004,fremat2005}. In this work we have corrected the measured $V\!\sin i$ quantities using similar curves to those given in \citet{fremat2005} but recalculated for the SMC metallicity. The corrected average rotational parameters per mass category are presented in Table~\ref{table1} as
$\langle V\!\sin i\rangle_{\rm corr}$.\par
\begin{figure}[h!t]
\centerline{\psfig{file=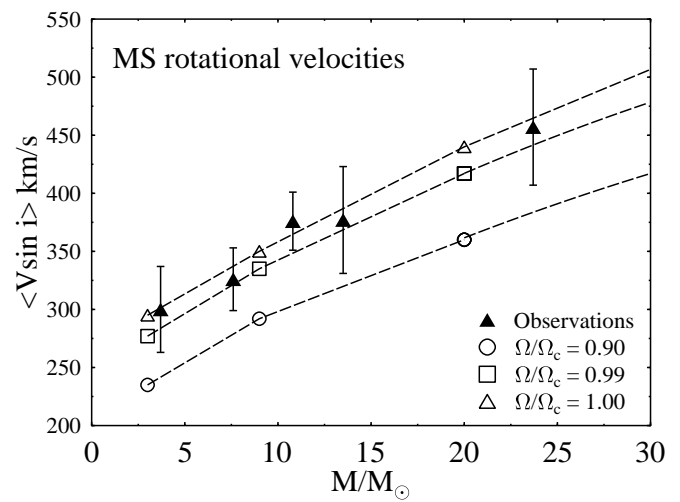}}
\caption{Comparison of the average observed rotational velocities SMC of Oe/Be stars with the locus of theoretical velocities for the SMC metallicity. The curves are for model stars that attain the surface angular velocity ratios \omc\,=0.90, \omc\,=0.99 and \omc\,=1.00, in the MS evolutionary phase.}
\label{fig1}       
\end{figure}

 
  Figure~\ref{fig1} shows the $\langle V\!\sin i\rangle_{\rm corr}$ per mass category of the studied SMC Oe/Be stars, compared with three new theoretical tracks of true equatorial rotational velocities calculated by \citet{ekstrom2008} for metallicity $Z\!=0.002$, which suits the SMC cluster NGC330 and its environment. In what follows, all models for SMC stars are for $Z\!=0.002$. To compare the theory with
observations, all model velocities were multiplied by $\pi/4=\langle\sin i\rangle$ to simulate the random distribution of the inclination $i$ of the rotational axis \citep{chandra1950}. Once the fundamental parameters of all studied stars had been determined, the models of rotating stars were chosen so as to match the average true rotational velocity (i.e. $V\!=\!(4/\pi)V\!\sin i$) per mass category at their respective average age. The average true rotational velocities, $V$, per mass category have uncertainties ranging from 20 to some 40 \kms (cf. Table~\ref{table1}). The inferred average ZAMS velocities for these mass categories are then affected by uncertainties that are of the same order of magnitude, i.e. $\delta V\sim\pm30$ \kms, on average \citep[][their Fig. 9]{marta2007}. For the near critical linear velocities, $V\!\sim V_{\rm c}$, and stellar masses $M\gtrsim17M_{\odot}$, the corresponding $|\delta V/V_{\rm c}|$  translates into $|\delta(\Omega/\Omega_{\rm c})|\lesssim0.05$ as the average uncertainty of the derived angle velocity ratios ($V_{\rm c}$ and $\Omega_{\rm c}$ are the linear and angular critical velocities, respectively).\par

 The obtained today $\Omega/\Omega_{\rm c}$ ratios for all mass groups are in the interval $0.99\lesssim\!\Omega/\Omega_{\rm c}\!\lesssim\!1.0$. Fig.~\ref{fig1} shows the loci of today $\langle V\!\sin i\rangle$ against the mass corresponding to $\Omega/\Omega_{\rm c}\!=0.99$ and 1.0. For comparison, this figure also includes the $\langle V\!\sin i\rangle$ curve for $\Omega/\Omega_{\rm c}\!=0.90$.\par 
 
  A similar comparison in \citet{marta2007}, made with observed rotational velocities that were not corrected for the underestimations induced by the GD effect, suggestes that the SMC Be stars rotate on average at \omc=0.95. Within the observational uncertainties, the corrected velocities for the GD effect shown in  Fig.~\ref{fig1}  likely correspond to \omc=0.99. We note that \omc=0.99 is for a linear velocity ratio $V/V_{\rm c}\!=\!0.97$ or for a ratio of surface centrifugal to gravitational accelerations $\eta\!=\!0.90$.\par

\begin{table}[h!t]
\caption[]{Values of the rotational velocities for Be and Oe stars for several average mass categories}
\centering
\begin{tabular}{cccccc}
\hline
\hline
\noalign{\smallskip}
Mass category                 & A   & B   & C    & D    & E    \\	
$\langle{M/M_{\odot}}\rangle$ & 3.7 & 7.6 & 10.8 & 13.5 & 23.7 \\
\noalign{\smallskip}
\hline	
\noalign{\smallskip}
N  & 14 & 82 & 13 & 15 & 7 \\
$\langle{V\!\sin i}\rangle$                   & 277 & 297 & 335 & 336 & 420 \\
\noalign{\smallskip}
$\langle{V\!\sin i}\rangle_{\rm corr}$        & 300 & 326 & 376 & 377 & 457 \\
\noalign{\smallskip}
$\log [{\rm age}({\rm yr})]$                  & 8.3 & 7.6 & 7.3 & 7.2 & 6.9 \\
\noalign{\smallskip}
$\langle{V}\rangle_{\rm ZAMS}^{\rm old}$      & 400 & 437 & 513 & 514 & $-$ \\
\noalign{\smallskip}
$\langle{V}\rangle_{\rm ZAMS}^{\rm new}$      & 448 & 487 & 562 & 563 & 683 \\
$\epsilon_{V_{ZAMS}}$                         &  35 &  27 &  23 &  23 &  44 \\
\noalign{\smallskip}
$\langle{V}\rangle_{\rm ZAMS}^{\rm magnet}$   & 389 & 423 & 488 & 489 & 643 \\
\noalign{\smallskip}
\hline
\noalign{\smallskip}
\multicolumn{6}{l}{N\,=\,number of stars by mass category}\\
\multicolumn{6}{l}{$\langle{V}\rangle_{\rm ZAMS}^{\rm old}$\,=\,average true rotational velocities in km/s}\\
\multicolumn{6}{l}{without correction for GD \citep{marta2007}}\\
\multicolumn{6}{l}{$\langle{V}\rangle_{\rm ZAMS}^{\rm new}$\,=\,average true rotational velocities in km/s}\\
\multicolumn{6}{l}{of the enlarged stellar sample, corrected for GD effect}\\
\multicolumn{6}{l}{$\epsilon_{V_{ZAMS}}$\,=\,error of $\langle{V}\rangle_{\rm ZAMS}^{\rm new}$ in km/s}\\
\multicolumn{6}{l}{$\langle{V}\rangle_{\rm ZAMS}^{\rm magnet}$\,=\,average true rotational velocities in km/s}\\
\multicolumn{6}{l}{of the enlarged stellar sample,}\\
\multicolumn{6}{l}{corrected for models including magnetic fields}\\
\noalign{\smallskip}
\hline
\noalign{\smallskip}
\end{tabular}
\label{table1}
\end{table}

\subsection{ZAMS rotational velocities}
\label{ZAMSrotvel}

\begin{figure}[t]
\centering{\psfig{file=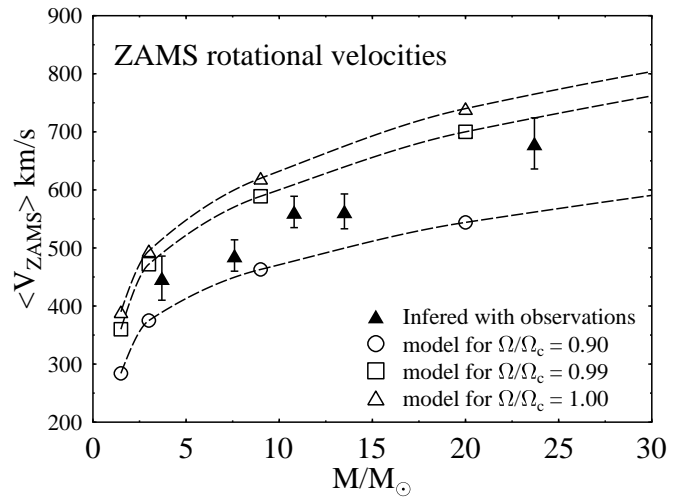}}
\caption{Inferred ZAMS rotational velocities of SMC Oe/Be stars (triangles) compared with model SMC ZAMS rotational velocities from \citet{ekstrom2008} (dashed curves).}
\label{fig2}       
\end{figure}

 Using the same approach as in \citet{marta2007}, which was based on models calculated by \citet{maeder01} and \citet{meynet02}, we re-determined the ZAMS rotational velocities of the SMC Oe/Be stars with a new and more complete grid of curves representing the evolution of rotational velocities at low metallicity
\citep{ekstrom2008}. The obtained results are given in Table~\ref{table1}. For the sake of comparison, this table also gives the previous average values of the ZAMS rotational velocities. The new results are shown in Fig.~\ref{fig2}, where the theoretical ZAMS rotational velocities calculated by \citet{ekstrom2008} are also drawn for the SMC metallicity and for the angular velocity ratios \omc\,=0.90,
\omc\,=0.99 and \omc\,=1.00. We notice then that SMC Oe/Be stars in the ZAMS rotate on average with \omc$\simeq\!0.95$. The models do not include magnetic fields, thus the mean rotational velocity on the MS is lower than it would be for magnetic models. We might thus overestimate the initial ratio \omc, which should be considered here as an upper limit. To take the difference in the models into account with magnetic fields, we compared the values of ZAMS velocities for given \omc\, values between the models from \citet{ekstrom2008} and from \citet{yoon2006}. The values of V$_{ZAMS}$ in the last row of
Table~\ref{table1} are then corrected by the corresponding scale factor.\par

  The trend of points inferred with the observation in Fig.~\ref{fig2} can be represented with $\langle V_{ZAMS}^{\rm obs}\rangle\!\simeq301\!\times\!M^{0.26\pm0.05}$ \kps. For the inference of the specific angular momentum that must have the circumstellar disks hanging over the collapsed stellar cores, it is instructive to estimate the average specific angular momentum, $\langle J/M\rangle_{ZAMS}$, that the SMC Oe/Be stars have in the ZAMS. To this end, let us write $\langle J/M\rangle_{ZAMS} = k_{\rm G}\langle V_{ZAMS}^{\rm obs}\rangle\langle R_{ZAMS}\rangle$, where $k_{\rm G}$ is the gyration radius and $R_{ZAMS}$ the stellar radius of stars in the ZAMS. The $k_{\rm G}$ and $R_{ZAMS}$ radii suited to \omc\,=\,0.95, chosen according to what is suggested by the observed points in Fig.~\ref{fig2}, varies with the mass as $k_{\rm G}\!=0.017\!\times\!M^{0.13\pm0.02}$, while the radius varies as $\langle R_{ZAMS}\rangle\!=1.72\!\times\!M^{0.59\pm0.02}$, \citep{zor1986,
zor1988,ekstrom2008}. By taking the decrease in the stellar moment of inertia induced by a rigid rotation at $\Omega/\Omega_{\rm c}\!=0.95$ into account, we obtain the following expression for $\langle J/M\rangle_{ZAMS}$, valid for the whole mass interval $3\lesssim M/M_{\odot}\lesssim30$; 

\begin{equation}
\langle J/M\rangle_{ZAMS} \simeq 6.0\pm0.1\times10^{16}\bigl(\frac{M}{M_{\odot}}\bigr)^{0.98\pm0.08}\  {\rm cm^2s^{-1}}\ ,
\label{eq1}
\end{equation}

\noindent which means that the average specific angular momentum of Oe/Be stars of the SMC in the ZAMS, considered as rigid rotators, increases linearly with the mass. Using moments of inertia calculated with stellar models at rest, \citet{kawaler1987} obtained for a mixed sample of B+Be stars in the MW that $\langle J/M\rangle\!\sim\!M^{1.43\pm0.16}$. Since the specific angular momentum of the last stable orbit around a Schwarzschild black hole scales with the mass $M/M_{\odot}$ as \citep{yoon2006} (3 times more than for a Kerr black hole)

\begin{equation}
(J/M)_{SBH} = 1.5\times10^{16}\bigl(\frac{M}{M_{\odot}}\bigr)\ \ {\rm cm^2s^{-1}} \ ,
\label{eq5}
\end{equation}

\noindent the SMC Oe/Be stars would then have a specific angular momentum only a few times greater than the one that is supposedly appropriate for producing a stable disk over the black hole after the stellar collapse. Since part of this angular momentum will be lost during the evolutionary following the MS phase, the SMC Oe/Be stars do not seem to have angular momenta that is to prevent the formation of circumstellar disks and/or the expected jets \citep{macfad1999}. \par 

 Because it may affect the properties of Oe/Be stars, after we take the average of the $\Omega/\Omega_{\rm c}$ rates over the masses in the ZAMS and in the MS, the equatorial velocity of these objects decreases from the ZAMS to the MS by a factor $0.76$, while $\Omega/\Omega_{\rm c}$ varies from \omc\,=\,0.95 to \omc\,=\,0.99. This change in $\Omega/\Omega_{\rm c}$ implies that the ratio 
$V/V_{c}$ varies from 0.87 to 0.97 and the ratio $\eta$ of the surface equatorial centrifugal to gravitational acceleration passes from $0.76$ to 0.90.\par

\section{LGRBs progenitors and the Oe/Be stars}
\label{lgrbp}

\subsection{ZAMS rotational velocities of LGRBs progenitors compared to those of Oe/Be stars}
\label{zrvp}

 Thanks to the combined effects mainly of rotation and metallicity, a series of models suggest that massive fast-rotating stars can fulfill the requirements imposed by the collapsar engine and be so progenitors of LGRBs \citep{hirschi2005, yoon2006}. To comply with the SN-LGRB relation, the involved stellar masses need to be higher than 10 M$_{\odot}$. Thus, while \citet{hirschi2005} conclude, on the basis of non magnetic models,  that for the SMC metallicity stars with masses between 32 and 92 M$_{\odot}$ are able to produce LGRBs, the latest models by \citet{yoon2006}, which include
magnetic fields decrease the lower limit of masses to those of early B-type stars. According to these models, fast rotation favors a strong mixing of chemicals, which means that the stars can evolve as quasi-chemically homogeneous and finally end up as massive helium stars. Since the lower the metallicity, the faster the stellar rotation can be, these phenomenon can in principle be favored in media with low metallicity, which accordingly impede strong stellar winds. In turn, this makes the stars conserve high amounts of angular momentum. The models by \citet{yoon2006}
predict a lower limit to the ratio of equatorial linear velocities $V/V_{\rm c}$ ($V_{\rm c}$ is the critical or breakup velocity) as a function of mass for the LGRB progenitors that do not seem to change strongly according to the metallicity, but the limiting $V/V_{\rm c}$ ratio increases if the stellar mass is lower. Moreover, the zone above the threshold $V/V_{\rm c}$ ratio of potential LGRB progenitors widens towards lower and higher masses as the metallicity decreases.\par 

 \citet[][ and references therein]{campana2008} indicate that the LGRB \object{GRB060218} had a progenitor with an initial mass of 20 M$_{\odot}$, i.e., a B0-B1 star, where the metallicity is $Z\!=\!0.004$, similar to the SMC. The chemical composition of the nebulae surrounding the star, if it is assumed that it reflects the chemical abundance of the progenitor, indicates  that the progenitor was a near critical rotator, as in the ``Be phenomenon''. Such a fast rotation is
required to produce an efficient mixing process in the star, which favors the stellar chemically homogeneous evolution and makes it possible for the environing nebulae reflect the stellar abundances.\par 
 
  To compare the models with the inferred ZAMS ratios \omc, we transformed the
`observed'\footnote{Actually, $(V/V_{\rm c})_{\rm ZAMS }$ must be understood as values inferred from observed $V/V_{\rm c}$ ratios through model predicted tracks of the stellar rotational velocity.} $(V/V_{\rm c})_{\rm ZAMS }$ ratios of the Oe/Be star subsamples in the SMC into \omc\ ratios by considering the $1\sigma$ dispersion of masses and equatorial velocities intervening in each average. The $V/V_{\rm c}$ and \omc\ ratios for massive stars are related as \citep{chauville01}

\begin{equation}
\frac{\Omega}{\Omega_{\rm c}} \simeq 1.381\frac{V}{V_{\rm c}}\bigl[1-0.276\bigl(\frac{V}{V_{\rm c}}\bigr)^2\bigr]\ ,
\end{equation}

\noindent which is also valid for the SMC metallicity. These results are given in Table~\ref{table2}. The indicated mass category, i.e. A$\pm\sigma$M, should be understood as $M({\rm A})\pm\sigma$M. The boldfaced value of the \omc\ in \% corresponds to the average mass $\langle{\rm M}\rangle$ of the corresponding subsample. The natural upper limit of the surface stellar rotation is obviously given by $V/V_{\rm c}\!=\!1.0$. In Table~\ref{table2} all values $V/V_{\rm c}\!>\!1.0$ are the
arithmetic consequences of the imposed symmetric $1\sigma$ dispersions and they should be understood as $V/V_{\rm c}\!\to\!1.0$. In Table~\ref{table3}, the values of Table~\ref{table2} are corrected for the effect of magnetic fields from the \citet{yoon2006} models as explained in Sect.~\ref{ZAMSrotvel}. \par

\begin{table}[h]
\caption[]{Ratios of the ZAMS \omc\ per mass category of SMC Oe/Be stars with the boldfaced values corresponding to the average \omc\ ratios of subsamples.}
\centering
\begin{tabular}{llll}
\hline
\hline	
\noalign{\smallskip}
Mass category   & \omc          & \omc$-\sigma$ & \omc$+\sigma$  \\
\noalign{\smallskip}
\hline	
\noalign{\smallskip}
A               &    {\bf 0.89}  &      0.79    &    1.01 \\
A$-\sigma$M     &      0.94      &      0.83    &    1.04 \\
A$+\sigma$M     &      0.85      &      0.75    &    0.96 \\
\noalign{\medskip}
B               &    {\bf 0.87}  &      0.68    &    1.08  \\
B$-\sigma$M     &      0.93      &      0.73    &    1.15  \\
B$+\sigma$M     &      0.82      &      0.64    &    1.01 \\
\noalign{\medskip}
C               &    {\bf 0.92}  &      0.71    &    1.08  \\
C$-\sigma$M     &      0.94      &      0.72    &    1.09  \\
C$+\sigma$M     &      0.91      &      0.70    &    1.06  \\
\noalign{\medskip}
D               &    {\bf 0.91}  &      0.63    &    1.07  \\
D$-\sigma$M     &      0.95      &      0.66    &    1.13  \\
D$+\sigma$M     &      0.87      &      0.60    &    1.02  \\
\noalign{\medskip}
E               &    {\bf 0.92}  &      0.78    &    1.02  \\
E$-\sigma$M     &      1.03      &      0.88    &    1.15  \\
E$+\sigma$M     &      0.82      &      0.70    &    0.91 \\
\noalign{\smallskip}
\hline
\end{tabular}
\label{table2}
\end{table}

\begin{table}[h]
\caption[]{Ratios of the ZAMS \omc\ per mass category of SMC Oe/Be stars corrected for the effect of magnetic field from the comparison between \citet{ekstrom2008} and \citet{yoon2006} models (boldfaced values correspond to the average \omc\ ratios of subsamples).}
\centering
\begin{tabular}{llll}
\hline
\hline	
\noalign{\smallskip}
Mass category   & \omc          & \omc$-\sigma$ & \omc$+\sigma$  \\
\noalign{\smallskip}
\hline	
\noalign{\smallskip}
A               &    {\bf 0.77}  &      0.69    &    0.88  \\
A$-\sigma$M     &         0.82   &      0.72    &    0.90  \\
A$+\sigma$M     &         0.74   &      0.65    &    0.83  \\
\noalign{\medskip}
B               &    {\bf 0.76}  &      0.59    &    0.94  \\
B$-\sigma$M     &         0.81   &      0.63    &    0.99  \\
B$+\sigma$M     &         0.71   &      0.56    &    0.88  \\
\noalign{\medskip}
C               &    {\bf 0.80}  &      0.62    &    0.94  \\
C$-\sigma$M     &         0.82   &      0.62    &    0.95  \\
C$+\sigma$M     &         0.79   &      0.61    &    0.92  \\
\noalign{\medskip}
D               &    {\bf 0.79}  &      0.55    &    0.93  \\
D$-\sigma$M     &         0.82   &      0.57    &    0.98  \\
D$+\sigma$M     &         0.76   &      0.52    &    0.89  \\
\noalign{\medskip}
E               &    {\bf 0.83}  &      0.70    &    0.92  \\
E$-\sigma$M     &         0.93   &      0.79    &    1.03  \\
E$+\sigma$M     &         0.75   &      0.63    &    0.82  \\
\noalign{\smallskip}
\hline
\end{tabular}
\label{table3}
\end{table}

\begin{figure}[t!]
\centering{\psfig{file=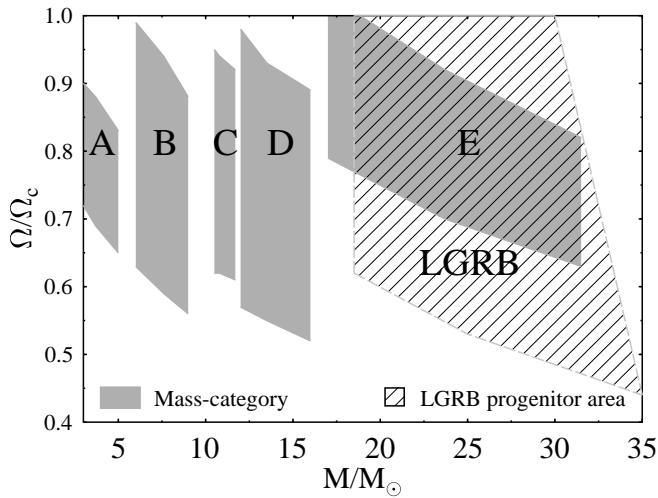}}
\caption{Comparison of the (\omc,~$M/M_{\odot}$) parameters in the ZAMS per mass category for SMC Oe/Be stars with the area of long gamma ray bursts progenitors at SMC metallicity predicted by \citet{yoon2006}.}
\label{fig3}       
\end{figure}

  The inferred ZAMS zones of $\Omega/\Omega_{\rm c}$ ratios per mass category of SMC Oe/Be stars given in Table~\ref{table3} are compared in Fig.~\ref{fig3} with the theoretically predicted region containing the LGRB progenitors for $Z\!=\!0.002$ by \citet{yoon2006}. One can see that only the more massive Oe/Be members, found in the ``E-category", may reach the zone of potential LGRBs progenitors. According to \citet{yoon2006}, the ratio of the core to envelope mass is low in the less-massive O/B stars, which favors an efficient braking of their core and thus a significant reduction of its specific angular momentum. These stars were then not included in our counting as potential LGRBs progenitors. Nevertheless, among the less massive OB objects, there may be many that are possibly storing high amounts of angular momentum, thanks to strong internal differential rotation. In these objects, the core specific angular momentum can then be higher than the one predicted with stellar evolutionary models constructed with an initial upper limit of kinetic energy imposed by the rigid rotation in the ZAMS. Thus, in spite of the limiting masses imposed by the above mentioned theoretical predictions, many more stars could be potential LGRBs progenitors. The same arguments can also be extended to stars, of the same and other masses, which are here excluded from the counting because of their apparent low surface rotation. We discuss this issue in Sect.~\ref{opprd}.\par 
  
 In Fig.~\ref{fig3} the shaded ``E" zone seems to show somewhat an unusual behavior compared to other mass categories. Although we are not able to specify which of them is dominant, four reasons may be used together to explain it: 1) the smallness of the stellar sample which by chance perhaps favors the highest $V\!\sin i$ values; 2) fast evolution of massive stars in low-metallicity regions that enables the initial high angular momentum to be retained; 3) the overestimated $V\!\sin i$ of massive Oe/Be stars because of some systematic non rotational line broadening effect, i.e. electron scattering, etc.\citep{ball95,zor92}; 4) the stronger mass-loss in the more massive stars, so that their surface $\Omega/\Omega_{\rm c}$ ratio is less likely to increase during the MS lifespan, see for examples \citet[][Fig. 11]{meynet2000} and \citet[][Fig. 4]{maeder01}.\par  
 
 \citet{hunter2008} indicated that, according to the \vsini~distributions of B
stars they observed in \object{NGC 346}, the areas of LGRBs progenitors could be more extended than expected from \citet{yoon2006} model predictions. The age of this cluster is roughly 3-5 Myrs \citep{nota2006,bouret2003}, there have been recent episodes of star formation and it hosted at least 1 supernova \citep{goul2008}. In this region there are also several classical Be stars and other emission-line objects \citep{wis2006,wis2007,hunter2008,martayan2009}.\par

\subsection{Evolution of Oe/Be and WR stars}

  Some stars could start in the ZAMS as born Oe/Be objects, but lose this property during their MS life-span due
to the spin-down produced by efficient mass-/angular momentum-loss phenomena. This can be the case of the massive
Be stars in the MW, which are only found in the first half of the MS \citep{zorec2005}. Contrary to the MW, the
massive Oe/Be stars and Oe stars in the SMC are also present in the second half of the MS. These stars can then
continue their post-MS steps by keeping fast rotational rates and evolve as chemically quasi-homogeneous objects
towards the fast-rotating WR phase \citep{yoon2006}. This prediction seems to be supported by
\citet{martins2009}, who find that several SMC WR stars are certainly fast rotating objects.\par 

  The finding in Sect.~\ref{zrvp} that the more massive members of the SMC Oe/Be population overlaps the LGRBs predicted progenitor area \citep{yoon2006}, thanks to their masses and $(\Omega/\Omega_{\rm c})_{\rm ZAMS}$ rotation rates, which are today in the second half of their MS evolutionary phase, makes them designed candidates for such chemically quasi-homogeneous objects that can evolve towards the WR phase and become LGRB progenitors later on.\par \citet{martayan2009} find that the ``Be phenomenon" can be shared by more numerous stellar populations, favored possibly by the lower SMC metallicity, than in the MW. Since the theoretical predictions show that the lower the metallicity the lower is the stellar mass that can enter the LGRB progenitor area, it is expected that among the massive stars of the first generations, which are very metal poor,  there might be a significantly large number of objects that could have undergone the ``Be phenomenon". These stars could then have had the required physical conditions for being potential LGRB progenitors.\par

\section{Predicted LGRBs rates from the Oe/Be star population and comparison with the observed ones}
\label{prates}

 Another test of the ability of the Oe/Be stars in the SMC to be LGRBs progenitors is
 based on the estimation of the expected rate of LGRB events that these stars can afford. In the less favorable case, the predicted frequency must be of the same order of
magnitude as the observed ones for this stellar population be considered a plausible
candidate.\par  

\subsection{The SMC Oe/Be stellar population and selection of candidates}
 To estimate the LGRB rate from the SMC Oe/Be stellar population, we need to identify the right spectral types to fill the mass requirements of stars entering the theoretically predicted LGRB progenitor zone shown in Fig.~\ref{fig3}. To this end, and knowing that we are dealing with fast rotators, we used the models of stellar evolution with rotation for stars with SMC metallicity calculated by \citet{maeder01}. We interpolated the evolution paths from the ZAMS to the TAMS for the masses $M\!=\!18M_{\odot}$ and $M\!=\!33M_{\odot}$. The $(\log L/L_{\odot},T_{\rm eff})$ corners thus obtained in the theoretical HR diagram were transformed into absolute magnitude-color pairs $[M_{\rm V},(V-I)_o]$ using standard bolometric corrections and the OGLE-III color-magnitude calibrations in current use. It follows from this that the main sequence visual absolute magnitudes of SMC stars with $M\!=\!18M_{\odot}$ range from $M_{\rm V}\!=\!-2.8$ mag to $M_{\rm V}\!=\!-4.4$, while those of $33M_{\odot}$ do it from $M_{\rm V}\!=\!-3.7$ mag to $M_{\rm V}\!=\!-6.3$. A mass-spectral type correspondence for stars in the ZAMS has also been recently obtained by \citet{huang2006}.\par
 
 In accordance with the MK spectral-types for dwarf stars in the SMC calibration against the visual absolute
magnitude $M_{\rm V}$ used in \citep{martayan2009}, the stars entering the LGRB progenitor zone in
Fig.~\ref{fig3} correspond then to those labeled with B1-O8 spectral types for rest and emission-less stars.
However, according to \citet{hunter2008} the mass of the LGRBs progenitors could be as low as 14M$_{\odot}$.  In
this case, one should count stars a little cooler than spectral type B1, i.e. from B1.5, up to O8 as potential
LGRBs progenitors with SMC metallicity. Stars from 14 to 33~$M_{\odot}$ evolve in the MS phase, so
that their respective absolute magnitudes $M_{\rm V}$ remain rather constant and that the $(V-I)_{0}$ color
changes by no more than $\delta(V-I)_{0}\sim0.04$ mag for 14~$M_{\odot}$ to $\delta(V-I)_{0}\sim0.07$ mag for
33~$M_{\odot}$. Moreover, in our [$M_{\rm V},(V-I)_{0}$] diagram, all counted OB stars are located in quite
a narrow strip of $M_{\rm V}$ as a function of $(V-I)_{0}$ \citep[see Fig. 5 in][]{martayan2009}, so that there cannot
be much confusion regarding the identification of stars by their masses with the employed photometric criterion.
\par

 The absolute magnitude intervals for 18 and $33M_{\odot}$ stars were established from evolutionary models where
the $(\log L/L_{\odot},T_{\rm eff})$ parameters are averaged over the rotationally deformed stellar surfaces.
Actual stars show, however apparent hemisphere-dependent spectroscopic and photometric characteristics. There are
then at least two reasons, rotation- and Be phenomenon-related, for which still cooler spectral types than B1
should enter the counting: 1) fast-rotating stars appear with cooler apparent spectral types than stars with the
same mass at rest \citep{fremat2005, marta2007}; 2) the largest number of candidate stars are selected using
photometric colors in the visible and near-IR, where both rotation and the presence of circumstellar disks make
the stars appear strongly reddened. We recall that reddening due to the circumstellar disks
$\Delta(B-V)\gtrsim0.05$ mag are currently observed in Oe/Be stars \citep{mouj1998,mouj1999,martayan2009}. These two effects
make that the stars with photometrically determined spectral types can appear from two to three subspectral
types cooler than expected for their intrinsic rest spectrum, so that in principle the counting could start at
spectral type B3.\par 

As a consequence of the many effects determining the apparent spectral type of Oe/Be stars,
the mass-spectral type mismatch can be high. Since the number of stars increases rapidly as the spectral type
becomes cooler, we prefer to deal with a lower limit of counted candidates. We thus considered stars only from
the apparent spectral type B2 and hotter. Once the number of all O/B stars, i.e. with and without emission lines,
that enter the chosen range of spectral types in the SMC, we determine the number of Oe/Be stars using the
frequencies of Be and Oe stars given by \citet{martayan2009}, which are based on a slitless survey of
emission-line stars in the SMC.\par 

\begin{figure}[t!]
\resizebox{\hsize}{!}{\includegraphics[angle=-90,width=7cm]{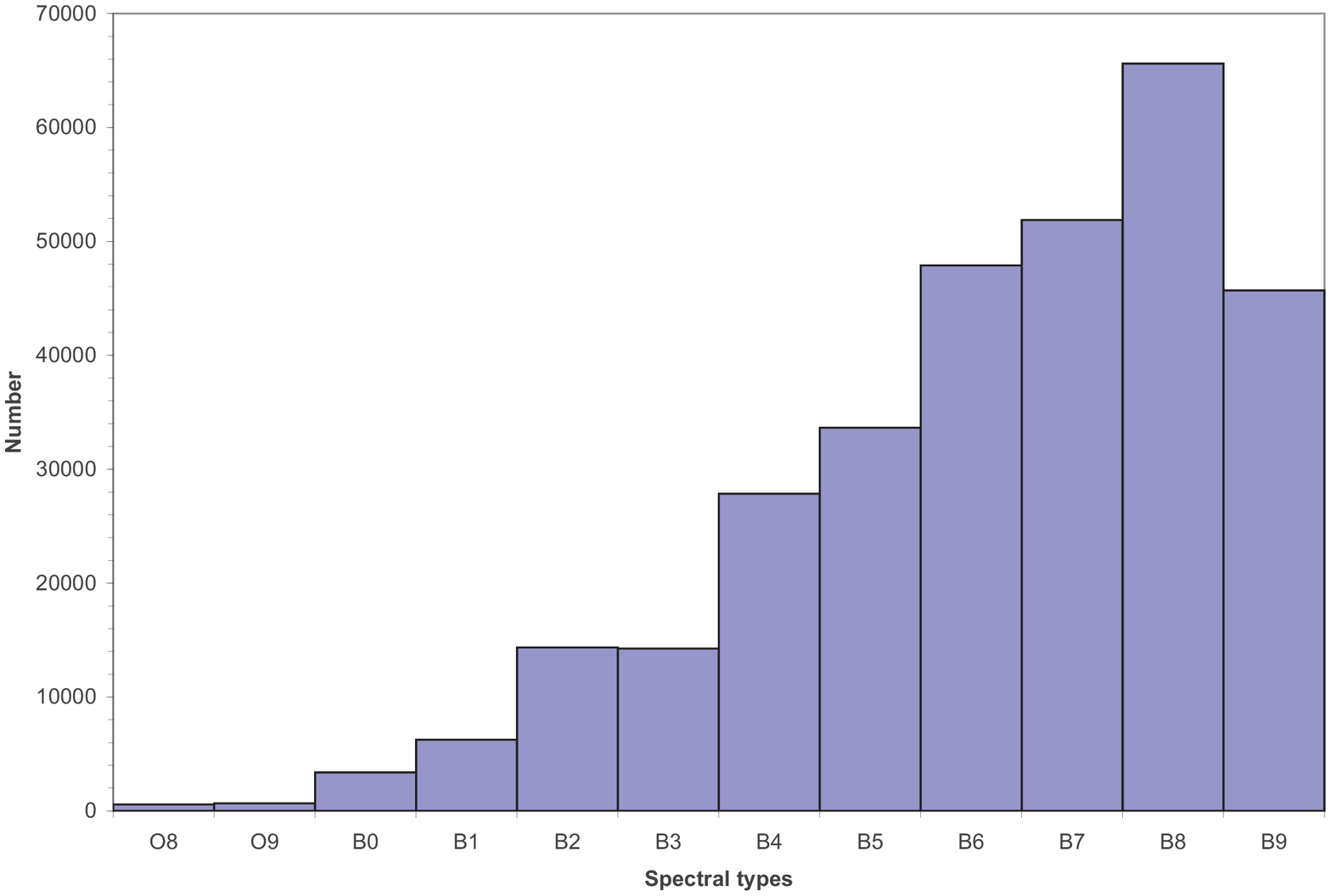}}
\caption{Spectral-types distribution of OB stars in the SMC counted with OGLE-III.}
\label{fig4}       
\end{figure}

 The counts of all O/B stars entering the desired spectral type ranges were performed using the new OGLE-III catalog \citep{Udalski2008}, which covers 14 square degrees, so that both the bar and the external fields of the SMC are covered. We note that OGLE-II data cover only 2.4 square degrees. Its use would require
extrapolating the number of OB stars for regions off the bar to infer the total number of O/B stars in the SMC with the required spectral types.\par

 The stars were selected using the dereddened color diagram [$M_{\rm V},(V-I)_o$]. To do the correction for the ISM extinction, we calculated the mean E[B-V] color excess based on the individual values from the SMC clusters \citep[see ][]{ageSMcocl}, which amounts to 0.084 mag and is similar to the 
E[B-V]=0.08 mag given by \cite{keller99b}. Then the absolute $M_{\rm V}$ and the dereddened $(V-I)_o$ were obtained following the same process as described in \cite{martayan2009}.
Using the calibrations by \cite{lang1992} and \cite{saga1997}, the E[V-I] is obtained by the formula:

\begin{equation}
E[V-I]= 1.44 \times E[B-V] + 0.0175 \times E[B-V]^{2}\ ,
\end{equation}
 where the second term is negligible with E[B-V]=0.084.
And finally the (V-I)$_{0}$
\begin{equation}
(V-I)_{0}= (V-I) -1.44 \times E[B-V]\ 
\end{equation}\par

 The number of OB stars in the SMC with and without emission lines so determined per subspectral type from B3 to O8 is shown in Fig.~\ref{fig4}, and detailed in Table~\ref{counts} for types between B3 and O8. The distribution in Fig.~\ref{fig4} shows that O and B-type stars basically follow the IMF except that the OGLE-III survey is not complete towards the late B stars, which explains the sudden decrease in counts at B9. We are finally left with the following reference number of objects to predict the rate of LGBRs events:\par

\begin{eqnarray}
\begin{array}{lcl}
N_{\rm tot} & = & \  4719\ {\rm stars\ hotter\ than\ B0} \\
N_{\rm tot} & = &  10988\ {\rm stars\ hotter\ than\ B1} \\
N_{\rm tot} & = &  25395\ {\rm stars\ hotter\ than\ B2} \\
N_{\rm tot} & = &  39726\ {\rm stars\ hotter\ than\ B3}. \\
\end{array}
\label{nbbe}
\end{eqnarray}

\begin{table}[]
\caption[]{Counts of OB main sequence stars with and without emission in the SMC per spectral type  with OGLE-III.}
\centering
\begin{tabular}{cccccc}
\hline
\hline	
\noalign{\smallskip}
O8  & O9  & B0   & B1   & B2   & B3  \\
\noalign{\smallskip}
\hline	
\noalign{\smallskip}
 566 & 671 & 3482 & 6269 & 14406 & 14332 \\
\noalign{\smallskip}
\hline
\end{tabular}
\label{counts}
\end{table}

 Since the FWHM of the OGLE-III PSF function is better than 1.34\arcsec, it implies that in the SMC the counted early-type stars must be separated by at least 0.33 pc for their images not to merge. According to a recent study by \citet{sabbi2009}, the highest early-type star density in the SMC is 8.5 star/pc$^2$, which implies mean stellar separations of 0.34 pc. This ensures that the counting in Table~\ref{counts} is complete.\par 
 
 In the next section we estimate the rates of LGRBs events and the number of LGRBs
in the local universe. Their purpose is to get the range of uncertainties that can affect the estimate of rates depending on those inherent to stellar counts and the angles of beams. \par 

\subsection{Prediction of the LGRB rates}
\label{fasm}

 To predict the number of LGRBs seen from Earth, we consider only specific beaming angles. In a first step, we
 determine the number of Oe/Be stars per spectral-type. We use for that the Be/(B+Be) and Oe/(O+Oe) ratios in the
 SMC determined by \cite{martayan2009}. The numbers of stars being sought are given in column 2 of
 Table~\ref{tableLGRBsmeth1}, where the ranges stand for the uncertainties on the Be/B fractions. The yearly
 frequency of LGRB events is then obtained by dividing the counts per mass category by the respective total
 lifetime. The stellar ages used were interpolated among those obtained for this purpose by \citet{yoon2006} for
 stars between 13 and 30M$_{\odot}$ with SMC metallicity. Assuming that the LGRBs only refer to single
 fast-rotating stars, we remove the binaries from the counts. As from \citet{porter2003} the binary rate is about
 30\% among Oe/Be stars, the obtained above rates must be multiplied by 0.7; however, the exact ratio of binaries
 in B-type stars is far from being well known. Among the hottest ones and in particular for O stars, the
 frequency of binary systems can reach 75\% \citep{sana2008,sana2009}. It is worth noting that only those
 binaries should be removed where the interaction between the components is significant, which is not always the
 case in Be-binary systems. Finally, we stress that the \citet{martayan2009} survey is a single-epoch survey.
 Because of the variable character of the Be phenomenon, according to \citet{fabregat03} and \citet{mcswain08}
 one third of the Be stars can be  missing. As a consequence, the final range of rates of LGRBs is calculated by
 multiplying the above rates by 0.7/(1-1/3). 
 
The LGRBs rates obtained for the SMC are called here the ``LGRB base-rates". They
represent the intrinsic rates that we generalize to all irregular galaxies of Magellanic type, which are metal poor and can host LGRBs progenitors. The LGRB base-rates are thus given by

\begin{equation}
R_{\rm LGRB}^{\rm br} = \left(\frac{0.7}{1-1/3}\right)\times\sum_{Sp}\left[\frac{N_{\rm Oe/Be}(Sp)}{t(Sp)}\right]\ \ \ {\rm LGRBs/yr}
\label{lgrbsbr}
\end{equation}

\noindent In this relation, $N_{\rm Oe/Be}(\rm Sp)$ is the number of stars with
spectral type $Sp$ displaying the Be phenomenon, and $t(Sp)$ is the total lifetime of stars
with spectral type $Sp$ in the early MS evolutionary phases. The constant factor in relation (\ref{lgrbsbr}) is of the order 1, which indicates a fortuitous compensation between the fraction of rejected stars for binarity and the completeness factor for missing Oe/Be stars. The $R_{\rm LGRB}^{\rm br}$ rates calculated with (\ref{lgrbsbr}) are given in column 3 of Table~\ref{tableLGRBsmeth1}.\par

 Since the LGRB phenomenon is strongly collimated, to obtain the potential
 rates that can be observed, in a second step we have to multiply the base-rates by the the probability that
 beaming angle sweep the Earth. The beam angles $\theta$ can range from 0.5$^{o}$ to 15$^{o}$
 \citep{hirschi2005,fryer2007,lamb2005,wats2006,zeh2006}. The probability that a beam of angle $\theta$ be seen
 at Earth is:

\begin{equation}
p(\theta) = \int_{0}^{\theta}\!\!\sin\theta\ {\rm d}\theta \ , 
\label{prob}
\end{equation} 

  Finally, to calculate the predicted yearly rate of LGRB events per average galaxy in a given volume of the Universe, we have to take the fraction of galaxies into account that have the required properties in this space  to host LGRB progenitors. Let us then assume that the SMC properties regarding the frequency of fast rotators with low metallicity is partaken by the irregular Magellanic-type galaxies (Im-type), which are metal poor. We note this fraction as $f_{\rm Im} = N_{\rm Im}/N_{\rm G}$, where $N_{\rm Im}$ is the number of Im-type galaxies in a given volume of the Universe, and $N_{\rm G}$ the number of all galaxies in the same volume of space. According to statistics by \citet{rocca2007}, it is $f_{\rm Im} = 0.17$, which is a very lower limit and valid up to at least redshift $z\sim0.5$.\par
  
 We present the results for the angles $\theta$ = : 5\degr, 10\degr, and 15\degr, which are used
by \cite{pod2004}. 
However, the angle $\theta$ = : 5\degr~according to \citet{wats2006} is quite frequent.
The angle 10$^o$ is roughly in the middle of the observed interval of opening angles and the angle 15$^o$ was chosen because is nearly the most opened one, although the least probable, but quite frequently evoked and used in the literature. 
It then happens that $p\!=\!0.0038$ for an angle of 5\degr, $p\!=\!0.015$ for 10\degr, and $p\!=\!0.0341$  for 15\degr. As there are 2 cones in each LGRB event, the probability (\ref{prob}) has to be multiplied by a factor 2. Thus, the predicted yearly rates of LGRB events per average galaxy produced in the local Universe, $R_{\rm LGRB}^{\rm pred}$, given in columns 4 to 6 of Table~\ref{tableLGRBsmeth1}
were obtained from

\begin{equation}
R_{\rm LGRB}^{\rm pred}(\theta) = R_{\rm LGRB}^{\rm br}\times2p(\theta)\times 
f_{\rm Im}.
\label{brpf}
\end{equation} 

\begin{table*}[ht]
\caption[]{LGRBs rates estimated with the first approach.}
\centering
\begin{tabular}{ccllll}
\hline
\hline	
\noalign{\smallskip}
Mass category   & Number of Be/Oe stars & LGRBs base rates & Proba 5\degr & Proba 10\degr &  Proba 15\degr \\
\noalign{\smallskip}
\hline	
\noalign{\smallskip}
B2e to O8e &  4978-6110  & 3.9-4.8 $\times 10^{-4}$ & 2.5-3.1 $\times 10^{-7}$ & 1.0-1.2 $\times 10^{-7}$ & 2.3-2.8 $\times 10^{-6}$ \\
B1e to O8e &  2774-3244  & 2.6-3.0 $\times 10^{-4}$ & 1.7-1.9 $\times 10^{-7}$ & 6.7-7.7 $\times 10^{-7}$ & 1.5-1.7 $\times 10^{-6}$ \\
B0e to O8e &  1483-1551  & 1.7-1.8 $\times 10^{-4}$ & 1.1-1.2 $\times 10^{-7}$ & 4.4-4.6 $\times 10^{-7}$ & 0.98-1.0 $\times 10^{-6}$ \\
O9e to O8e &   257-294   & 5.3-6.2 $\times 10^{-5}$ & 3.4-4.0 $\times 10^{-8}$ & 1.4-1.6 $\times 10^{-7}$ & 3.1-3.6 $\times 10^{-7}$ \\
     O8e   &   118-135   & 2.4-2.8 $\times 10^{-5}$ & 1.6-1.8 $\times 10^{-8}$ & 6.2-7.2 $\times 10^{-8}$ & 1.4-1.6 $\times 10^{-7}$ \\
\noalign{\smallskip}
\hline
\end{tabular}
\label{tableLGRBsmeth1}
\end{table*}

 The observed yearly rate of LGRB events per average galaxy that we use as a reference in this work to which we compare the predicted ones from OeBe stars is obtained from the total rate of GRBs, of which 2/3 are considered of long-duration. From the BATSE monitoring it is found that $R^{\rm obs}_{\rm LGRB}\!\sim\!(0.2-3)\times10^{-7}$ LGRBs/year/galaxy, where the extreme values concern the local Universe or the whole Hubble volume, respectively \citep{zhme2004,pod2004,fryer2007}. According to the $(\Omega/\Omega_{\rm c};M/M_{\odot})$-area of LGRBs progenitors indicated by \citet{yoon2006} and the mass-calibration by \citet{huang2006} or \citet{lang1992}, but neglecting spectral type changes from rotational and circumstellar effects discussed in Sect.~\ref{prates}, we should only consider counts of stars hotter than spectral types B0-B1. Taking the beaming angles from $5^o$ to $15^o$ into account, we note that the estimates made with populations of Be/Oe stars from B0e to O8e overlap the  observed range of LGRBs rates adopted above. The hottest populations (O9e to O8e) alone are not able to reproduce the rates.\par

 Another way of comparing the predicted rates with the observed ones is to determine the number of LGRBs observed in the local Universe up to a given redshift z. The number of LGRB events that have taken place in the local Universe and have been 
seen from Earth during the last $Y$ years, is given by:

\begin{equation} 
N_{\rm LGRB}^{\rm pred} = R_{\rm LGRB}^{\rm pred}(\theta)\times N_{\rm G}\times Y,
\label{nlgrbs}
\end{equation}

\noindent where $N_{\rm G}$ is the number of galaxies of all type counted up to a redshift
$z$. We chose $z =0.2$, since the number of galaxies in this volume seems to be about well determined. From the 2MASS catalog \citep{Skrutskie2006}\footnote{see also http://www.haydenplanetarium.org/universe/  duguide/exgg\_twomass.php}, we obtain 
$N_{\rm G}(z\!\leq\!0.2)=$ 1\,140\,931 galaxies. Adopting $Y=11$ years (see below), the predicted number of LGRB events
seen from Earth are given in Table~\ref{tableLGRBlocal}.\par

 Using the GRBox from the University of California at Berkeley\footnote{see http://lyra.berkeley.edu/grbox/grbox.php?starttime  =670702\&endtime=091231} for years between 1998 and 2008, there are 8 LGRBs (080108, 060614, 060505, 060218, 051109B, 031203, 030329A, 980425) and 3 SGRBs (061201, 050709, 000607) with a redshift lower than 0.2. This proportion of 73\% of LGRB and 27\% of SGRB agrees with the proportion of 75-25\% as reported by \citet{zhme2004} for the LGRBs-SGRBs. 

\begin{table*}[ht]
\caption[]{Number of LGRBs predicted in the local universe (z$\le$0.2) in 11 years.}
\centering
\begin{tabular}{lccc}
\hline
\hline	
\noalign{\smallskip}
Mass category   & Number for angle=5\degr & Number for angle=10\degr & Number for angle=15\degr \\
\hline	
\noalign{\smallskip}
B2e to O8e &  3-4  & 11-14  & 25-31   \\
B1e to O8e &  2-2  & 7-9    & 16-19   \\
B0e to O8e &  1-1  & 5-5    & 11-11   \\
O9e to O8e &  0-0  & 2-2    & 3-4     \\
     O8e   &  0-0  & 1-1    & 2-2     \\
\noalign{\smallskip}
\hline
\end{tabular}
\label{tableLGRBlocal}
\end{table*}

\begin{figure}[t!]
\centering{\psfig{file=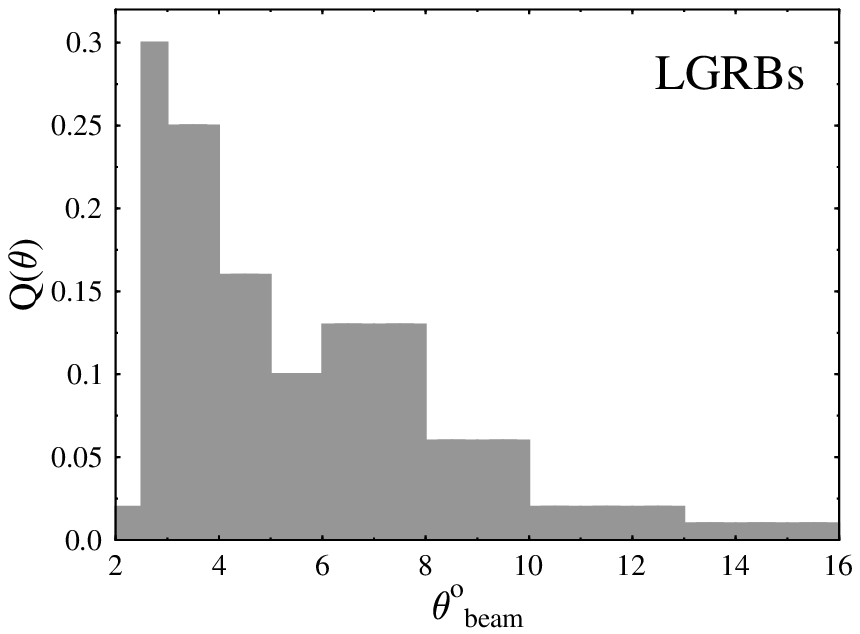}}
\caption{Frequency of beam angles presented by \citet{wats2006} and \citet{zeh2006}.}
\label{fig5}       
\end{figure}

  \citet{zeh2006} and \citet{wats2006} have studied the distribution of opening angles, which is reproduced in
Fig.~\ref{fig5}. If the LGRB event rate was estimated considering this distribution of beam angles, the
$2p(\theta)$ probability factor in relation (\ref{brpf}) had to be replaced by the integrated probability $P$
given by

\begin{equation}
P = 2\times\!\int_{0^o}^{16^o}\!\!p(\theta)Q(\theta){\rm d}\theta = 0.015 \ , 
\label{cone}
\end{equation}

\noindent where $p(\theta)\!=\!1-\cos\theta$ is given by (\ref{prob}), and $Q(\theta)$ is the distribution shown
in Fig.~\ref{fig5}. From (\ref{cone}) we deduce that $P$ defines an average opening angle of $7^o$, roughly. We
can then interpolate the corresponding LGRB rate or number of events in Tables 5 and 6, respectively, which gives
$N_{\rm LGRB}^{\rm pred}\!\sim\!(2-5)\times10^{-7}$ LGRBs/yr/galaxy or $N_{\rm LGRB}^{\rm pred}\!\sim\!3-6$ LGRB
events in the past 11 years. On account of the uncertainties that may affect not only our knowledge of the
opening angle of beams or their distribution, but also the estimated rates of observed LGRB events in the local
and Hubble Universe, we may consider that the counts limited to the spectral type interval B0-B1 to O8 predict
rates that overlap rather well with the observed ones. This is the first time that an estimate of the LGRBs rates
is done on the basis of a well-defined population of stars, i.e., born fast rotators with low initial
metallicity.\par

 Very recently, \citet{georgy2009} has tried to reproduce the observed rates of LGRBs with models of WR stars. Although they obtained higher values of LGRBs rates than what observations seem to indicate, these authors conclude that only a fraction of WC/WO stars among those that initially were rapid rotators could produce gamma ray events. This fast-rotation argument may then focus some interest on stars with the characteristics discussed in the present work.\par
 
 The counts of Oe/Be candidates for LGRB progenitors, as they were done here, however, represents a mere lower limit. Stars with masses corresponding to rest B0-B1 spectral types that rotate at $\Omega/\Omega_{\rm c}\gtrsim0.7$ following the requirements shown in Fig.~\ref{fig3} then display apparent spectral types cooler by at least one subspectral type. A prudent lower limit set at spectral type B2, would nevertheless enhance our predictions by a rough factor 2, as seen from counts displayed in (\ref{nbbe}), or Tables~\ref{tableLGRBsmeth1} and \ref{tableLGRBlocal}. However, massive fast rotators without emission lines should also be considered potential LGRBs progenitors as we see in Sect.~\ref{frwel}, so that the final predicted rates may widely surpass the observed ones.\par

\section{Other phenomena and possible biases in the LGRBs rate determination}
\label{opprd}

\subsection{Binaries}
\label{binar}

 If the LGRBs could also occur in a binary system \citep{cantiello2007, vdh2007} under the same conditions as fast rotation and chemical composition, the rates we estimated above should be multiplied by 1/0.7.
However, the results from \citet{langer2006} suggest that the binary evolutionary path is unlikely to produce LGRBs when they are associated with metal-poor galaxies \citep{kewley2007}.\par 

\subsection{Fast rotators without emission lines}
\label{frwel}

 The ZAMS rotational velocities of Oe/Be stars in the MW form the tail of a true rotational velocity distribution $\Phi(V)$ that starts at a velocity limit given by $V_{\rm L}\!=\!30\times(M/M_{\odot})$ \citep{zorec2007}. The tail of the velocity distribution formed by the Oe/Be stars can be represented with a Gaussian function, whose dispersion is also mass-dependent and is roughly given by $\sigma_{\rm V}\!=\!28\times(M/M_{\odot})$ \citep{zorec2007}. This implies that there are rotators that
can only acquire the Be phenomenon at the very end of their MS evolutionary phase, which cannot be seen as Oe/Be stars today. On the other hand, there are also fast rotating stars in the ZAMS that have slightly lower rotational velocities than $V_{\rm L}$ required to become Oe/Be, which will never display emission lines. However, it can easily be shown that they have a global specific angular momentum that is equal to or
larger than given by Eq.~\ref{eq5}, so that they certainly have enough angular momentum to be considered as candidates for LGRBs progenitors.\par 

Let us also take into account those objects which in principle have the required angular momentum but were not included in our counting to derive the LGRBs rate. To this end we assume that they are rigid rotators. The equatorial velocity of all those having the specific angular momentum given by Eq.~\ref{eq5} is readily calculated to give $V_{\rm J}\!=\!100\times(M/M_{\odot})^{0.28}$. The fraction of objects yet not included in the estimated LGRB rates is then represented by the correcting factor $q$ calculated as

\begin{equation}
q = \frac{1-{\rm erf}(x_{\rm L})}{1-{\rm erf}(x_{\rm SBH})}\ ,
\label{eq4}
\end{equation}

\noindent for which, as noted above, we have taken the high velocity tail $\Phi(V)$ given by a Gaussian function. In Eq.~\ref{eq4} ${\rm erf}(x)$ is the error function, $x_{\rm L}\!=\!V_{\rm L}/\sqrt{2}\sigma_{\rm V}$, and $x_{\rm SBH}\!=\!V_{\rm J}/\sqrt{2}\sigma_{\rm V}$. The factor $q$ then ranges from 1.2 to 1.5 for masses from 17 to 30$M_{\odot}$. We can then multiply the LGRB rates found in Sect.~\ref{prates} by an average factor $\overline{q}\!\simeq\!1.35$, which increases even more the predicted rate well above the observed one and makes the massive fast-rotating stars highly probable
progenitors of LGRBs.\par 

Finally, from the equatorial velocity $V_{\rm J}$ required to attain the stellar specific angular momentum of the order given by Eq.~\ref{eq5}, we derive the angular velocity ratios that range from $\Omega/\Omega_{\rm c}\!=\!0.50$ to 0.52 for stars in the E-category in the ZAMS. This means that the LGRBs progenitor area: E mass-category of Fig.~\ref{fig3}, can be entirely covered by the O/B-type fast rotators. Assuming that the distribution of the rotational velocities of the stars entering the LGRBs progenitor region in Fig.~\ref{fig3} foreseen by \citet{yoon2006} is the same as used to calculated the fraction $q$ in relation (\ref{eq4}), it follows that our E mass group represents 44\% of the stellar population in the LGRBs region.\par  

\subsection{Differential rotators}

  Models of stars that start in the ZAMS as rigid rotators, transform their original internal uniform angular
velocity into a profile of differential rotation, where the stellar core rotates faster than the envelope
\citet{meynet2000}, in no more than 2-3\% of their MS lifespan. The latest theoretical works of stellar evolution
with rotation claim, however, that magnetic fields through the Tayler-Spruit instability
\citep{spruit1999,spruit2002,maeder2005,yoon2006} are able to maintain the rigid rotation. Nevertheless, a recent
critical discussion of the efficiency of the Tayler-Spruit instability by \citet{zahn2007} shows that the action
of magnetic fields can be much less efficient than previously thought. In all works of stellar structure, it is
also assumed that convection imposes $\Omega\!=$ constant. \par

A number of discussions show that convection likely redistributes the specific angular momentum $j\!=\Omega r^2$
rather than $\Omega$, which implies that in the convection zones the angular velocity should be $\Omega\propto
r^{-p}$, where $p$ is a positive quantity \citep{tayler1973,deupree1998}. In the Sun, the differential
rotation exists precisely in the convective zones! Since the stars do not start their life in the ZAMS, but have
passed though a long story before this phase, their internal rotation could already be differential in the ZAMS
and with a larger content of angular momentum than just the allowed by a profile of rigid rotation. For an
average star of 20$M_{\odot}$, it is then a simple matter to calculate what must be in order of magnitude its ZAMS
ratio of energies $\tau\!=\!K/|W|$, higher than some maximum $\tau_{\rm crit,rigid}\!=\!0.015$ allowed by the
rigid rotation ($K$ is the rotational kinetic energy; $W$ is the gravitational potential energy) in order for the
star to keep up the same appearance as a rigid rotator throughout its MS evolutionary span. For that, we
simply ask that the star fulfill at least two requirements: a) at any Lagrangian mass-coordinate the star never
reaches the fateful limit $\tau\!\simeq\!0.14$, which can make it secularly unstable and would tend to transform
the delimited stellar volume into a three-axial Jacobian ellipsoid; b) during the whole MS phase it has on the
surface $V_{\rm equator}\lesssim V_{\rm critical}$. Calculations show that these conditions can be fulfilled by
$\tau_{\rm ZAMS}\!\simeq\!0.03$, which is roughly two times $\tau_{\rm crit,rigid}$ in the ZAMS. \par

If by chance O/B stars started their MS life span as neat differential rotators with $\tau\!>\!\tau_{\rm
crit,rigid}$, the internal mixing processes can be strong in a large number of stars where this phenomenon is
unexpected a priori because of their apparently slow rotation. Then make their chemical composition attain the
required properties to become potential LGRBs progenitors. Consequently, the estimate of the SMC $R_{\rm LGRB}$
rate we made in this paper might still be increased with stars of a wider variety of masses and apparent low
surface velocities.\par 

\subsection{Underluminous GRBs}
 
 Recently, it has been discovered that underluminous GRBs \citep{fryer2007,foley2008} could not be observed by the gamma ray satellites or not be classified as a GRB. According to these authors, this finding indicates that there may be a population of events that are less luminous but are possibly from 10 to $10^2$ times more frequent. If these events are also considered, the local rate of GRBs could then amount to $\sim10^{-6}-10^{-5}$ LGRBs/year/galaxy. If correct, this last rate will indicate that other masses or kinds of stars can also be concerned such as less-massive B stars, in particular Bn stars. these stars are objects with spectral types that are most frequently later than B7. Moreover, they are fast rotators that do not have emission lines.\par

\subsection{Other possible progenitors}

 According to \citet{hammer2006}, runaway WR stars could be another potential progenitor, but it is very difficult to estimate their impact on the LGRBs rates.\par

\subsection{The ``first stars'' and  the LGRBs}

 Using the GRBox from 1998 to 2009, we computed the statistics of GRBs by type with intermediate (z$\ge$3) and high redshifts (z$\ge$5). In the intermediate-z category, most of the GRBs are found to be long-GRBs (29 LGRBs, 2 SGRBs, 8 unknown). In the high-z category, the result is similar, 6 are LGRBs or type II, 2 are unknown, and 0 are SGRBs. The very high redshift GRB080913 was reclassified as a type II GRB by \cite{zhang2009}. And the most distant object known, GRB 090423, which has a redshift of 8.2, is also an LGRB or type II GRB \citep{salvaterra2009,tanvir2009, zhang2009}. According to \cite{salvaterra2009}, the properties of the very-high redshift LGRBs are similar to those of low/intermediate redshift. According to \citet{yoon2006} models, the number of LGRBs should increase with the redshift and the decrease in Z until the first stars. However, the number of LGRBs, which can originate in Pop III stars, is limited by this type of population. At high redshift it is expected that the stars have very low metallicity and  are similar to the first stars. In this case, they can reach rotational velocities as high as 800 \kms, or even higher, as shown by \cite{chiappini2006} and \cite{ekstrom2008b}. \cite{ekstrom2008b} and \cite{meynet2007} show that these stars can also retain enough angular momentum in their core to collapse into black holes and produce an LGRB. Moreover, \cite{hirschi2007} show that the stars with extremely low metallicity can rotate very fast and produce LGRBs. All these theoretical results agree with the statistics reported above and indicate that the stellar rotation close to the breakup velocity, as in the Be-phenomenon at low metallicity, is a key ingredient for the production of LGRBs. As a consequence, the LGRBs at high redshift should find their origin in the fast-rotating first stars.

\section{Conclusion}

 While the first models focused primarily on fast-rotating WR stars as probable progenitors LGRBs, the latest developments indicate that fast-rotating, low-metallicity massive B- and O-type stars could also be behind the LGRBs. Such massive B- and O-type stars can actually be the Be and Oe stars of low metallicity found in the SMC. To test this assumption, we used fundamental parameters derived for Oe/Be stars in the SMC observed with VLT-FLAMES instruments. We first compared the average \vsini~determined for several mass subsamples of SMC Oe/Be stars with the theoretically predicted rotational velocities. We found that today these objects on average have a surface angular velocity ratio \omc$\simeq0.99$. Then, we determined the ZAMS rotational velocities again for Be stars in the SMC that we had studied in previous works and determined the ZAMS rotational velocities for a new subsample composed by SMC Oe stars. The average angular velocity rate of all these objects in the ZAMS is $\langle$\omc$\rangle\!=\!0.95$. The ZAMS \omc~rates thus determined were compared with those of fast-rotating massive stars foreseen by model calculations of LGRBs progenitors having an initial metallicity similar to the average one in the SMC. This comparison indicates that massive Oe/Be stars in the SMC could certainly have the properties required to be taken as LGRBs progenitors.  A lower limit of 17\% of galaxies in the near Universe are of irregular Magellanic-type that, by definition, all have metallicities $Z\lesssim0.002-0.004$. All can then host LGRBs progenitors. From the discussion it appears that not only can massive Oe/Be stars be LGRBs progenitors, but a larger population of O/B stars without emission, which are less massive and with apparent lower surface rotation velocities, can also partake of this quality. The total number of LGRBs estimated from the Oe/Be star populations based on spotted beaming angles and counts from B0-B1 to O8 stars is on average $R_{\rm LGRB}^{\rm pred}\!\sim\!10^{-7}$ to  $\sim\!10^{-6}$
LGRBs/yr/galaxy, which represent $N_{\rm LGRB}^{\rm pred}\!\sim\!2$ to 14 in 11 years.
These predictions  increase beyond the observed ones, if the stellar counting starts at spectral types B1-B2, in accordance with the apparent spectral types of massive fast rotators. Massive fast rotators without emission lines can still enhance these predictions by more than 30\%. On account of all possible uncertainties that may affect the calculated and the observed rates of LGRBs in the local Universe, we can consider that both overlap correctly. Young star formation regions in the SMC like NGC346 could then have hosted or will host one LGRB in the next Myrs.\par

\begin{acknowledgements}
 The authors acknowledge the referee R. Hirschi for his very useful comments, that helped to clarify and improve this article. The GRBOX is created and maintained by Daniel Perley (UC, Berkeley).  C.M. thanks C. Ledoux for useful information about the under luminous GRBs. This research has made use of the Simbad and Vizier databases maintained at the CDS, Strasbourg, France, of NASA's Astrophysics Data System
Bibliographic Services, and of the NASA/IPAC Infrared Science Archive, which is operated by the Jet Propulsion Laboratory, California Institute of Technology, under contract with the US National Aeronautics and Space Administration. This publication makes use of data products from the Two Micron All Sky Survey, which is a joint project of the University of Massachusetts and the Infrared Processing and Analysis Center/California Institute of Technology, funded by the National Aeronautics and Space Administration and the National Science Foundation.
\end{acknowledgements}

\bibliographystyle{aa}
 \bibliography{13079bib}

\end{document}